\documentclass{tMPH2e}
\usepackage{amsmath,amssymb} 
\usepackage{graphicx,psfrag}
\begin{document}
\doi{10.1080/0266476YYxxxxxxxx}
\issn{1360-0532}
\issnp{0233-1934}
\jvol{00} \jnum{00} \jyear{2014} \jmonth{January}

\markboth{Taylor \& Francis and I.T. Consultant}{Molecular Physics}


\title{Dynamics of a monodisperse Lennard-Jones system on a sphere}

\author{Julien-Piera Vest$^{\ast}$\thanks{$^\ast$Corresponding author. Email: vest@lptmc.jussieu.fr
\vspace{6pt}}, Gilles Tarjus and Pascal Viot$^{\rm b}$\\\vspace{6pt}  {\em{ Laboratoire de Physique Th\'eorique de la Mati\`ere Condens\'ee, UPMC, CNRS  UMR 7600,
 4, place Jussieu, 75252 Paris Cedex 05, France}};  }
\maketitle

\begin{abstract}
We investigate by  Molecular Dynamics simulation a system of $N$ particles moving on the surface of a two-dimensional 
sphere and interacting by a Lennard-Jones potential. We detail the way to account for the changes brought by a nonzero curvature, both 
at a methodological and at a physical level. When compared to a two-dimensional Lennard-Jones liquid on the Euclidean plane, where 
a phase transition to an ordered hexagonal phase takes place,  we find that the presence of excess defects imposed by the topology of the 
sphere frustrates the hexagonal order. We observe at high density a rapid increase of the relaxation time when
the temperature is decreased, whereas in the same range of temperature the pair correlation function of the system 
evolves only moderately.
\end{abstract}
\bigskip

\section{Introduction}

There are physical situations in which particles are irreversibly adsorbed to a substrate with a nonzero curvature and remain mobile 
on this substrate so that dense equilibrated phases can be formed.  
This may take place for instance on the wall of a porous material or the surface of a large solid particle\cite{Vincent1980,Hansen1980}, or else at an interface in an oil-water 
emulsion\cite{Aveyard2003,Tarimala2004,Bausch2003,Lipowsky2005,Irvine2010,Leunissen20022007,Irvine2012}. In such cases, the fluid is confined to a curved 
two-dimensional manifold. A simple example is that of a liquid layer on the surface 
of a spherical drop or particle. The curvature of the substrate, which is positive and constant for the sphere, influences the phase behavior, 
the structure and the dynamics of the adsorbate. 
In the cases under consideration, the geometry is frozen, providing a background whose metric and topological characteristics 
affect the behavior of the embedded fluids and solids, but with no feedback influence from the latter.

In addition to its relevance to physical problems, studying the properties of a system in curved space provides an extra control parameter 
(the Gaussian curvature) on top of the common thermodynamic parameters. This may prove interesting in several 
situations. For instance, in computer studies of models in the Euclidean space with the presence of long-range interactions 
and/or long-range spatial correlations between particles, boundary conditions matter and may lead to a subtle handling of the  
thermodynamic limit\cite{Kratky1980,Caillol1991,Hansen1979,Caillol1992,Caillol1999}. A possible alternative is then to implement instead 
``spherical boundary conditions'', which amount to considering 
the system on the surface of a sphere (or a hypersphere). The system is now of finite extent but does not require boundary conditions 
as it is compact. The thermodynamic limit can then be achieved by increasing the radius of the sphere or hypersphere (then decreasing 
the curvature), which may prove more convenient than using boundary conditions directly in Euclidean space.\cite{Moore_Perez-Garrido_1998,Moore_Perez-Garrido_1999}.
Another 
application of curved space is that it allows one to thwart crystallization of simple fluids in two dimensions\cite{Nelson2002,Bowick2001}. The hexagonal ordering which is 
prevalent in the Euclidean plane is indeed frustrated by curvature which introduces an irreducible density of topological defects and then 
opens the possibility of observing glassy behavior\cite{PhysRevB.62.8738,Nelson2002,Tarjus2005,Sausset2008}.

In this paper, we study by Molecular Dynamics simulation a monodisperse fluid of atoms confined on a surface of constant positive 
curvature, namely the surface of a sphere $S^2$, and we mainly focus on the dynamics of the system as a function of temperature. Some 
time ago, we investigated the same system on a two-dimensional manifold of constant negative curvature, namely the hyperbolic plane 
$H^2$\cite{Sausset2008,Sausset2010,Sausset2010a}. 
Besides the sign of the curvature, the differences between $S^2$ and $H^2$ are that the former is of finite extent whereas the latter can be 
infinite and that contrary to $S^2$, $H^2$ cannot be embedded in three-dimensional 
Euclidean space and therefore lacks a direct physical realization\cite{Tarjus2011}. 
The comparison of the effects of positive versus negative curvature is nonetheless 
an interesting prospect.

\section{Model and simulation method}
We study the monodisperse Lennard-Jones model embedded in the $2$-dimensional surface of a sphere $S_2$. 
The pair interaction potential is
\begin{equation}
\label{interactionLJ}
 v( r)=4\epsilon\left[\left(\frac{\sigma}{r}\right)^{12}-\left(\frac{\sigma}{r}\right)^6\right],
\end{equation}
where $r$ is the geodesic distance between two atom centers on $S_2$ and the interaction is truncated at a conventional cutoff 
distance of $2.5 \sigma$. This choice of interaction depending on the geodesic distance is also known as the ``curved line of force''\cite{Post1986}. The units of 
mass, length, energy and time are $m$, $\sigma$, $\epsilon$, and $\sqrt{m\sigma^2/\epsilon}$. 
Molecular Dynamics simulations are done in the micro-canonical ensemble with constant number of particles $N$, volume $V$ (actually, surface area), and energy $E$. 
Due to the spherical geometry, fixing $V$ selects the radius $R$ of the sphere $S^2$. We have chosen a reduced density $\rho=8 R^2[1-\cos(\frac{\sigma}{2R})](N/V)=
(2N/\pi)[1-\cos(\frac{\sigma}{2R})]=0.92$. (In flat space, \textit{i.e.} when $R\rightarrow \infty$, $\rho=(N/V)\sigma^2$, as commonly used.)
The relevant dimensionless quantity measuring the relative radius of curvature is $2R/\sigma$, which at fixed $\rho$ and $R$ (or $V$) is then 
determined by the number of particles $N$. In the present work, we have considered the case $N=1000$, which corresponds to 
$2R/\sigma\approx 18.6$.

One first derives the equations of motion. The  Lagrangian reads 
\begin{equation}
 L=\sum_{i=1}^N \frac{m{\bf v}_i^2}{2}-\frac{1}{2}\sum_{i\neq j}{v(r_{ij})}.
\end{equation}
where ${\bf v}_i$ is the velocity of the particle $i$ and $r_{ij}$ is the geodesic distance between particles $i$ and $j$,
\begin{equation}
r_{ij}=R\arccos\left(\frac{{\bf r}_i.{\bf r}_j}{R^2}\right).
\label{eq:20}
\end{equation}

Since particles are moving on the sphere, one has to add  the holonomic constraints $f_i({\bf r}_i)={\bf r}_i^2-R^2=0$ for $i=1\cdots N$.
By introducing $N$ Lagrangian multipliers $\alpha_i$, one defines  an extended Lagrangian\cite{Ryckaert1977,Leeuw1990} as 
\begin{equation}
 L^*=\sum_{i=1}^N \frac{m{\bf v}_i^2}{2}-\frac{1}{2}\sum_{i\neq j}{v(r_{ij})}+\sum_{i=1}^N \alpha_i f_i({\bf r}_i).
\end{equation}
The Lagrange equations $\frac{d}{dt}\left(\frac{\partial L^*}{\partial {\bf v}_i }\right)=\frac{\partial L^*}{\partial {\bf r}_i}$ then give 
\begin{equation}\label{eq:lag}
m\ddot{{\bf v}}_i=2\alpha_i{\bf r}_i-\sum_{j\neq i}\frac{\partial v(r_{ij})}{\partial {\bf r}_i}.
\end{equation}
After taking the scalar product of Eq.(\ref{eq:lag}) by ${\bf r}_i$ and using the second time derivative of $f({\bf r}_i)$, one obtains an explicit 
solution for the Lagrange multiplier,
\begin{equation}\label{eq:lag2}
 \alpha_i=\frac{1}{2}\sum_{j\neq i}\frac{\partial v(r_{ij})}{\partial {\bf r}_i}.{\bf r}_i-m\frac{{\bf v}_i^2}{2R^2}.
\end{equation}
By using the vector identity of the double cross-product and inserting  Eq.~(\ref{eq:lag2}) in Eq.~(\ref{eq:lag}), one finally obtains
\begin{equation}
m\dot{{\bf v}}_i=\frac{1}{R^2}\left[-m{\bf v}_i^2{\bf r}_i+{\bf r}_i\times \left({\bf r}_i\times \sum_{j\neq i}
\frac{\partial v(r_{ij})}{\partial {\bf r}_i(t)} \right)\right].
\end{equation}

It is convenient to express the equations of motions in term of positions and angular velocities instead of  positions and velocities.
Denoting  by $\boldsymbol{\omega}_i$ the  angular velocity of the  particle $i$, one has   
\begin{equation}\label{eq:lag3}
{{\bf v}_i} =\boldsymbol{\omega}_i \times {\bf r}_i  
\end{equation}
and the equations of motion become
\begin{equation}
 m\dot{{\boldsymbol \omega}}_i=-{\bf r}_i\times \sum_{j\neq i}\frac{\partial  v(r_{ij})}{\partial {\bf r}_i(t)}.
\end{equation}

One can write a ``velocity Verlet algorithm'' as follows. Let $\Delta t$ the time step. The angular velocity is updated as
\begin{equation}
 {\boldsymbol \omega}_i (t+\Delta t)= {\boldsymbol \omega}_i (t)-\frac{\Delta t}{2m}\left({\bf r}_i (t+\Delta t)
 \times \sum_{j\neq i}
 \frac{\partial  v(r_{ij})}{\partial {\bf r}_i}(t+\Delta t)+{\bf r}_i (t)\times \sum_{j\neq i}
 \frac{\partial  v(r_{ij})}{\partial {\bf r}_i}(t)\right).
\end{equation}
Through  Eq.~(\ref{eq:lag3}) the  position update  is then given by
\begin{equation}
 {\bf r}_i (t+\Delta t)= [1+a(t,\Delta t)] {\bf r}_i (t)+ \Delta t\left({\boldsymbol \omega}_i (t)+\frac{\Delta t}{2}\dot{{\boldsymbol \omega}}_i\right)\times {\bf r}_i,
\end{equation}
where $a(t,\Delta t)$ is determined by enforcing  the constraints  ${\bf r}_i^2 (t+\Delta t)={\bf r}_i^2 (t)=R^2$\cite{Ryckaert1977,Lee_Leok_2009}.

One finally obtains that 
\begin{align}
{\bf r}_i(t+\Delta t)&=\left(1-||\Delta t{\boldsymbol \omega}_i(t)-\frac{\Delta t^2}{2m} {\bf r}_i(t)\times
\sum_{j\neq i}\frac{\partial  v(r_{ij})}{\partial {\bf r}_i}
(t)||^2\right)^{1/2}{\bf r}_i(t)\nonumber\\
&
+\left(\Delta t{\boldsymbol \omega}_i (t)-\frac{\Delta t^2}{2m} {\bf r}_i(t)\times 
\sum_{j\neq i}\frac{\partial  v(r_{ij})}{\partial {\bf r}_i}
(t)\right) \times {\bf r}_i(t).
\label{eq:24}
\end{align}

For the initial configuration, particles are placed randomly on the sphere with the constraint that the distance between any pair of particles 
is always larger than $0.85\sigma$. In the early stage of the simulation, the velocities are rescaled in order that the mean 
kinetic energy becomes equal to a given temperature $T$. 

\section{Measured quantities}

To characterize the structure of the system, we have computed the pair correlation function
$g(r)$, where $r$ is the geodesic distance between particle centers on the surface of the sphere (and is limited to $\pi R$ 
due to the finiteness of the latter). The function $g(r)$ on $S^2$ is computed as in the Euclidean space from the density of particle centers 
at a distance between $r$ and $r+dr$ of a given particle  and averaged over all the atoms of the system.

We have also studied the topological defects, disclinations and dislocations, which are defined with respect to an underlying hexagonal order. 
In two dimensions, these defects are point-like and can be defined at a microscopic level through a Voronoi construction or its dual, 
the Delaunay triangulation. These constructions allow one to uniquely determine the number of nearest neighbors of any given 
atom. A hexagonal or hexatic environment corresponds to $6$ nearest neighbors and disclinations, which break bond-orientational order, 
appear as atoms with strictly less (positive disclination) or more (negative disclination) than 6 nearest neighbors. On the other hand, 
dislocations, which break translational order, appear as ``neutral dipoles'' formed by a positive and a negative disclination. 
The Delaunay triangulation of the atomic configurations on $S^2$ has been implemented with the Stripack library\cite{Renka1997}. 
The computation of the triangulation scales as $O(N\ln(N))$.

Defects are quite generally induced by temperature, but in curved space, there is an additional source of them. The topology of the 
embedding space indeed constrains the density of defects. In two dimensions, this is immediately seen from the Euler-Poincar\'e 
relation\cite{PhysRevB.62.8738,Bowick2001,Tarjus2011},
\begin{equation}
\frac{N}{6}(6-\overline{z})=\chi,
\label{eq_euler-poincare}
\end{equation}
where $\overline z$ is the mean coordination number of the atoms and $\chi$ is the Euler characteristic of the surface, which 
is equal to 0 for the Euclidean plane and 2 for the sphere $S^2$. As a result, there must be an excess of atoms 
with less than 6 neighbors, \textit{i.e.} of positive disclinations, in the latter. The minimum number of such defects is 12 disclinations of 
``topological charge" $+1$ (12 atoms with 5 neighbors) in an otherwise 6-fold coordinated medium, which then fulfills 
the constraint of Eq. (\ref{eq_euler-poincare})\cite{PhysRevB.62.8738}.

To describe the dynamics, we have studied the mean distance traveled by the atoms and the self-intermediate scattering 
function. In Euclidean space, the latter is simply the Fourier 
transform of the self-part of the density-density time correlation function,
\begin{equation}
\begin{aligned}
F_s(\mathbf{k},t)&=\frac{1}{N}{\sum_{j=1}^{N}{<\exp(i\mathbf{k}(\mathbf{r}_j(t)-\mathbf{r}_j(0))>}}
\\&=\frac{1}{N}{\sum_{j=1}^N{<\cos(k d_j(0,t))>}},
\end{aligned}
\end{equation}
where $d_j(0,t)$ is the distance traveled by atom $j$ during times $0$ and $t$. The generalization of this definition to spherical 
geometry is based on the appropriate extension of the Fourier transform and reads\cite{Tarjus2011}
\begin{equation}
\begin{aligned}
F_s(k,t)=\frac{1}{N}{\sum_{j=1}^N{<P_{kR}\left(\cos\left(\frac{d_j(0,t)}{R}\right) \right)>}},
\end{aligned}
\end{equation}
where $kR$ is an integer, $P_n$ is the $n$-th Legendre polynomial, and $d_j(0,t)$ the geodesic distance traveled by atom $j$ between $0$ and $t$ 
on the surface of a sphere of radius $R$.

As we are interested by the structural relaxation associated with a local probe on the scale of the interatomic distance, we have taken  
$k$ as the integer part of $\pi/\sigma$ and we have extracted the relaxation time $\tau$ as the time at which $F_s(k,t)$ reaches $0.1$.

\section{Results}

\begin{figure}[h!]
\begin{center}
 \includegraphics[width=10cm]{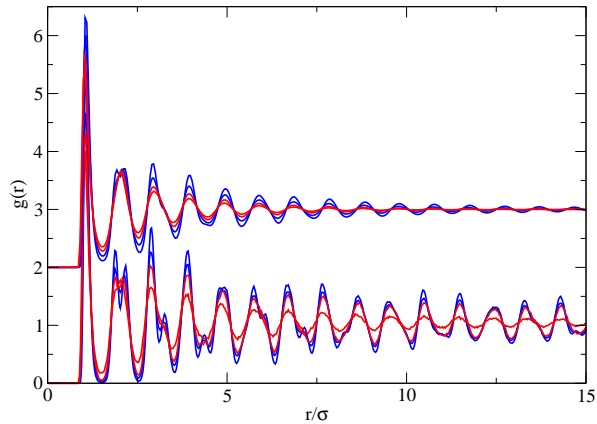}
  \caption{Pair correlation function $g(r)$ as a function of the (geodesic) distance $r$  for a Lennard-Jones system of 
  $1000$ particles at a reduced density $\rho=0.92$ and for temperatures  $T$
 from $2$ (red)  down to  $0.8$ (blue).  The top curves are for the sphere $S^2$ (results  shifted by a step of $2$) and 
the bottom  curves are for  the Euclidean plane $E^2$ with periodic boundary conditions.}\label{fig:gder}
\end{center}
\end{figure}

The pair correlation function $g(r)$ is displayed for several temperatures in Fig.~\ref{fig:gder}. We show the result for the Lennard-Jones system 
both on the sphere $S^2$ with $N=1000$ atoms (top) and on the Euclidean plane $E^2$ (bottom). 
The liquid becomes more structured as temperature decreases but on $S^2$ it never develops sustained oscillations at long range, which would be 
characteristic of true crystalline order. This is to be contrasted with the behavior of  $g( r)$ on the Euclidean plane, for which oscillations at low temperature are 
much more pronounced and signal the establishment of a crystalline-like phase.

\begin{figure}
\begin{minipage}{100mm}
\subfigure[]{
\resizebox*{5cm}{!}{\includegraphics{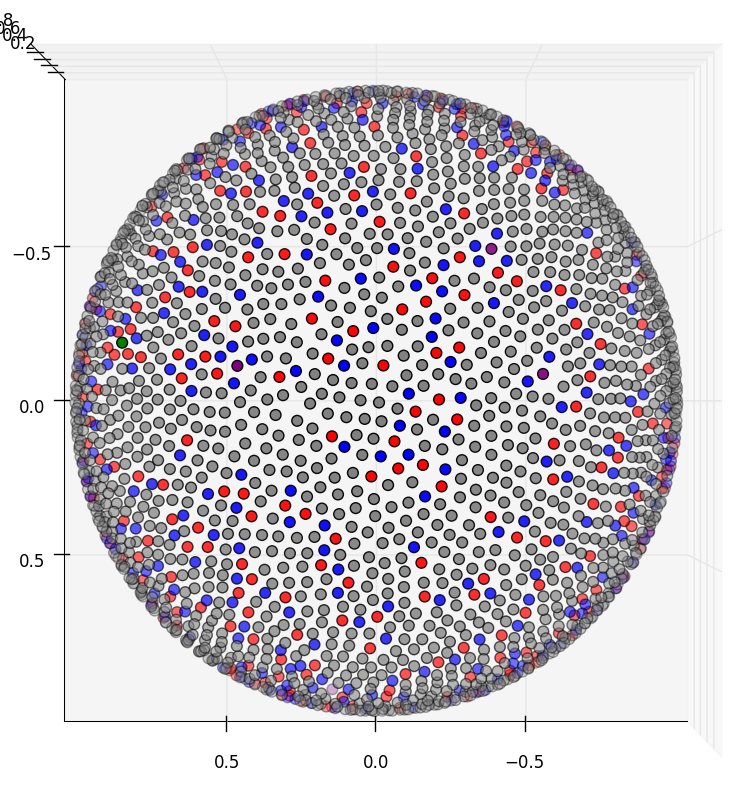}}}%
\subfigure[]{
\resizebox*{5cm}{!}{\includegraphics{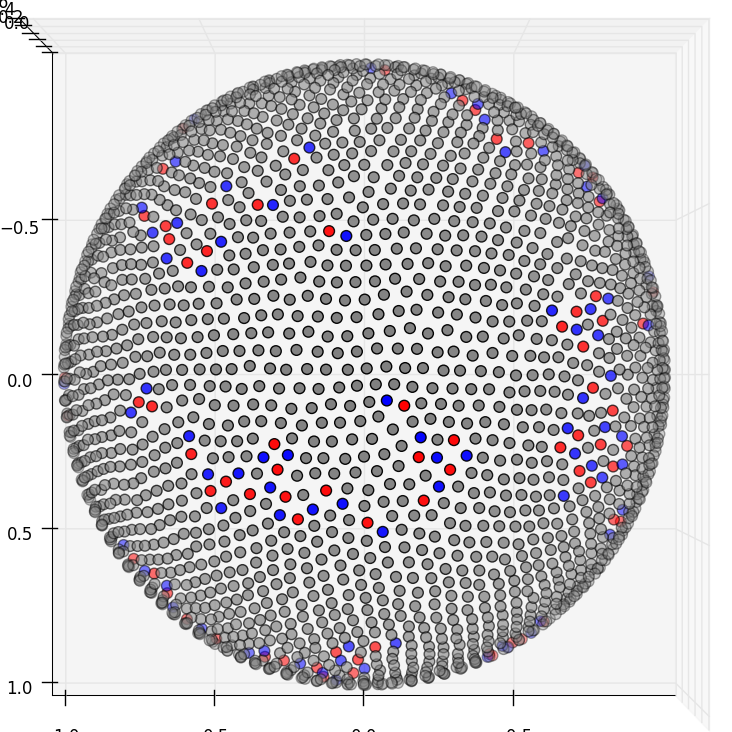}}}%
\subfigure[]{
\resizebox*{5cm}{!}{\includegraphics{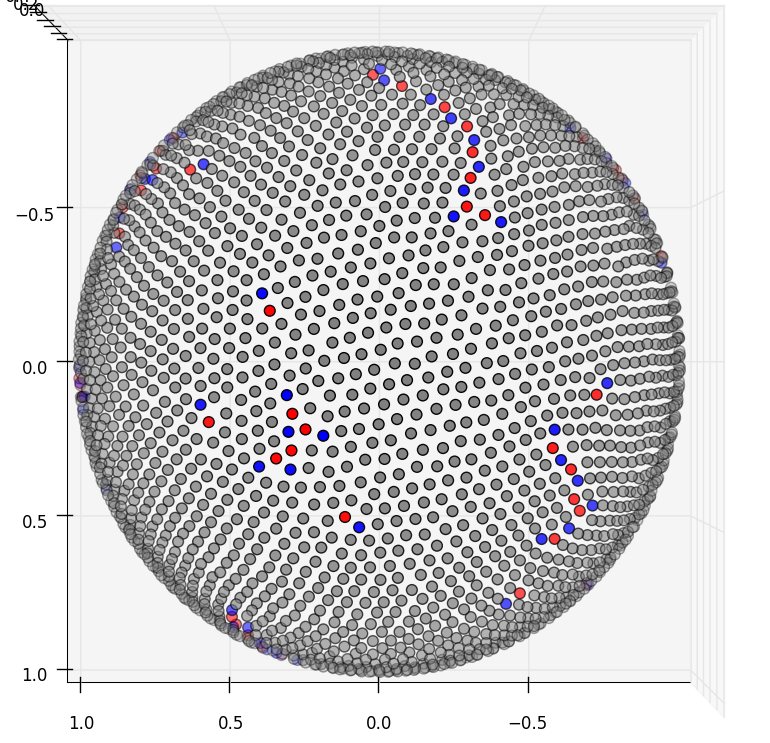}}}%
\end{minipage}
\caption{Typical atomic configurations at different temperatures: (a) $T=2$, (b) $T=1$ and (c) $T=0.5$. Red particles correspond to $7$ 
neighbors (negative disclination), blue particles to $5$ particles (positive disclination) and grey particles to $6$ neighbors.}%
\label{fig:snap}
\end{figure}

The absence of perfect long-range order can be visualized by looking at typical atomic configurations with a color code to highlight the topological 
defects. This is shown in Fig.~\ref{fig:snap} where gray atoms correspond to a 6-fold (hexatic or hexagonal) environment, while red and blue atoms 
respectively correspond to 5-fold and 7-fold environments, \textit{i.e.} to positive and negative disclinations of unit topological charge. At high 
temperature, above the freezing temperature in the Euclidean plane, many defects are present and scattered all over the system (Fig.2a). As one cools 
the system, these thermal defects progressively disappear (Fig.2b) and at low temperature one essentially observes strings of defects, which have been 
called grain ``boundary scars''\cite{PhysRevB.62.8738} and experimentally observed \cite{Bausch2003,Irvine2012}.
There are exactly 12 of them, which are formed of dislocations (red-blue pairs) attached to an excess positive (red) disclination. At the temperature $T=0.4$ (Fig. 2c), 
there are still a few isolated thermal dislocations that should disappear at still lower temperature. These defects, disclinations, dislocations and grain boundary 
scars disrupt the hexatic and hexagonal order of the system, which explains the form of the pair correlation function discussed above.

\begin{figure}[h!]
\begin{center}
  \resizebox*{11cm}{!}{ \includegraphics{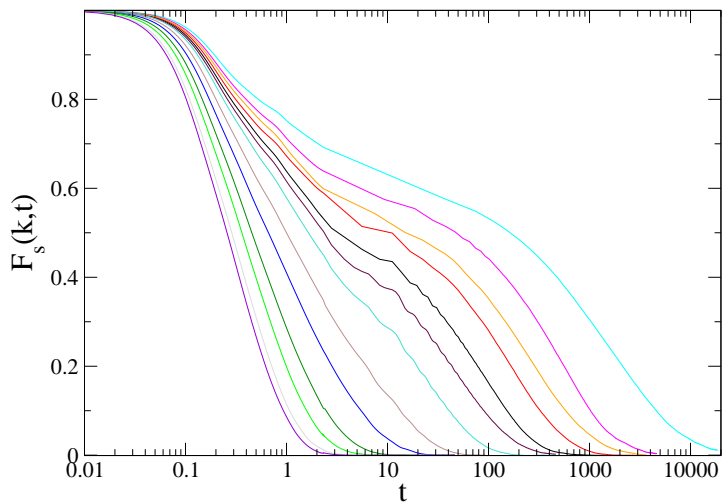}}
  \caption{ Log-linear plot of the self-intermediate scattering function $F_s(k,t)$ of a Lennard-Jones system of 
  $1000$ particles on $S^2$ at a reduced density $\rho=0.92$ for various temperatures 
  $T=3,2.5,2,1.6,1.2,1,0.8,0.7,0.6,0.55,0.5,0.4$ (from left to right). }\label{fig:Fs}
\end{center}
\end{figure}

\begin{figure}[h!]
\begin{center}
  \resizebox*{11cm}{!}{ \includegraphics{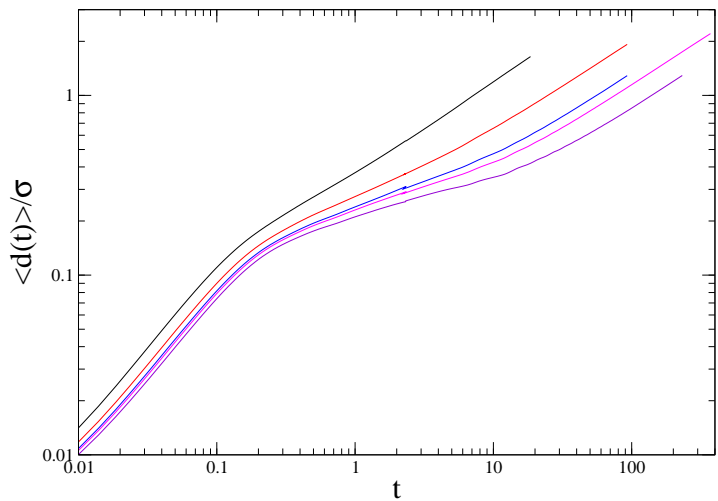}}
  \caption{Mean distance traveled by an atom versus  time for a Lennard-Jones system of 
  $1000$ particles on $S^2$ at a reduced density $\rho=0.92$ and various temperatures 
 $T=2,1.6,1.2,0.9,0.8$ (from top to bottom). }\label{fig:displacement}
\end{center}
\end{figure}

\begin{figure}[h!]
\begin{center}
  \resizebox*{11cm}{!}{ \includegraphics{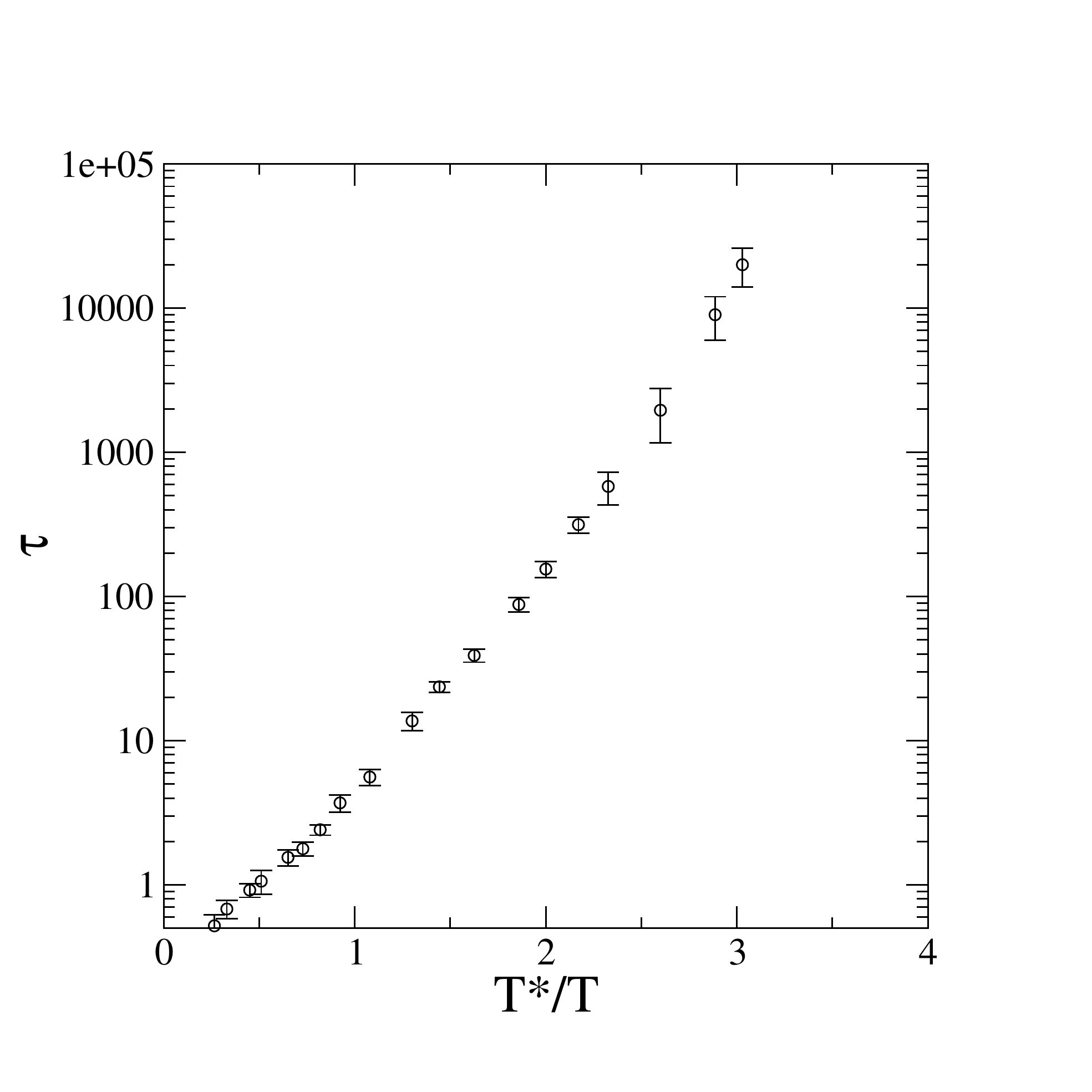}}
  \caption{Arrhenius plot of the relaxation time $\tau$ versus the reduced inverse temperature $T^*/T$, where $T^*\approx 1.3$ 
  is an estimate of the temperature at which ordering in a hexagonal phase takes place in the Euclidean plane.  The parameters are 
  the same as in Fig.~\ref{fig:Fs}.}
  \label{fig:tho}
\end{center}
\end{figure}

Turning now to the dynamics, we display the self-intermediate scattering function $F_s(k,t)$  in Fig. \ref{fig:Fs} and the mean traveled distance 
$<d(t)>$  in Fig. \ref{fig:displacement}, both 
being plotted versus the logarithm of the time and for different temperatures. A clear slowing down of the dynamics is seen as one cools the 
system: this is illustrated in Fig. \ref{fig:tho} where we display  the evolution of the relaxation time on an Arrhenius plot. 
For temperatures below the freezing temperature in the Euclidean plane $T^*\approx 1.3$, a shoulder develops in the self-intermediate scattering function 
and in the mean traveled distance.This shoulder indicates that relaxation begins to proceed in two regimes separated by a growing plateau region 
where not much relaxation takes place. Such a feature is characteristic of ``glassy'' dynamics and should be further investigated along the lines of 
previous work in the hyperbolic plane\cite{Sausset2008,Sausset2010a}.

\section{Conclusion}

We have investigated the structure and the dynamics of a simple atomic fluid confined to the surface of a sphere.
Such a system is relevant both for describing experimental realizations of colloidal particles adsorbed on a spherical 
substrate and for studying the role of curvature-induced frustration on the slowing down of the dynamics in liquids. 
For this work, we have developed a Molecular Dynamics algorithm that allows one to generate particle 
trajectories in a spherical geometry and we have adapted the tools familiar to studies 
of liquids in Euclidean space to curved space. The first results show that cooling the system slows down the 
dynamics with features that resemble those seen in glass-forming liquids and that in place 
of crystallization to a hexagonally ordered phase (technically, a quasi-long 
ranged ordered phase, in two dimensions) the system forms at low temperature patches of hexagonal order disrupted
by strings of topological defects known as grain boundary scars. Future work will involve 
investigating the effect of the curvature of the 
substrate, \textit{i.e.} the radius of the sphere compared to the radius of the particle, on the structure
and the dynamics of the liquid.

\appendix
\section{Spherical coordinates}
By using the spherical coordinates, the number of degrees of freedom for a point on $S^2$ 
matches the number of coordinates. The velocity of particle $i$ has two nonzero components: 
$v_\theta=R\dot{\theta}, v_\phi=R\sin(\theta)\dot{\phi}$.
The Lagrangian of the system is then given by 
\begin{equation}
L=\sum_{i=1}^{N}{\frac{1}{2}m R^2(\dot{\theta_i}^2+\sin^2\theta_i\dot{\phi_i}^2)}-\frac{1}{2}\sum_{i\neq j}{v(r_{ij})},
\label{eq:3}
\end{equation}
where $r_{ij}$, which is the distance between particle $i$ at position $(\theta_i,\phi_i)$ and particle $j$ 
at position $(\theta_j,\phi_j)$, is given by 
\begin{equation}
r_{ij}=R\arccos(\cos\theta_i\cos\theta_j+\sin\theta_i\sin\theta_j\cos(\phi_i-\phi_j)).
\label{eq:2}
\end{equation}

The Lagrange equations read
\begin{eqnarray}
mR^2\frac{d^2\theta_i}{dt^2}=mR^2\sin(\theta_i)\cos(\theta_i)\left(\frac{d\phi_i}{dt}\right)^2+F_{\theta_i}(t),\\
\label{eq:9}
mR^2\left(\sin(\theta_i)\frac{d^2\phi_i}{dt^2}+2\cos(\theta_i)\frac{d\phi_i}{dt}\frac{d\theta_i}{dt}\right)=F_{\phi_i}(t),
\label{eq:10}
\end{eqnarray}
with
\begin{eqnarray}
F_{\theta_i}(t)&=&\sum_{j\neq i}{A(r_{ij})[\sin(\theta_i)cos(\theta_j)-\cos(\theta_i)\sin(\theta_j)\cos(\phi_i-\phi_j)]},\\
\label{eq:11}
F_{\phi_i}(t)&=&\sum_{j\neq i}{A(r_{ij})\sin(\theta_j)\sin(\phi_i-\phi_j)},
\label{eq:12}
\end{eqnarray}
and
\begin{equation}
A(r_{ij})=\frac{24\epsilon}{\sigma}\left[2\left(\frac{\sigma}{r_{ij}}\right)^{13}-\left(\frac{\sigma}{r_{ij}}\right)^7\right]\frac{R}{\sin(r_{ij}/R)}.
\label{eq:13}
\end{equation}
We note that Ref.\cite{Moore_Perez-Garrido_1998} used a Verlet velocity algorithm in the framework of the spherical coordinates on $S^2$. 
However, we found that it is unstable for large system sizes and long times.
\\

We would like to dedicate this paper to our colleague and friend Pierre Turq for this special issue celebrating his contribution to Statistical Mechanics.
We thank F. Sausset for useful discussions and input.

\label{lastpage}

\end{document}